\documentclass[conference]{IEEEtran}
\usepackage[export]{adjustbox}

\usepackage{cite}

\usepackage{graphicx}
\usepackage{epstopdf}
\usepackage{amsmath}
\usepackage{bm}
\usepackage{amssymb}
\usepackage{mathrsfs}
\usepackage{subfigure}
\usepackage{url}
\makeatletter

\newcommand{\Rmnum}[1]{\expandafter\@slowromancap\romannumeral #1@}
\makeatother

\usepackage{algorithmic}
\usepackage{algorithm}
\ifCLASSOPTIONcompsoc
  \usepackage[caption=false,font=normalsize,labelfont=sf,textfont=sf]{subfig}
\else
  \usepackage[caption=false,font=footnotesize]{subfig}
\fi
\begin{document}
%
\title{Bayesian Design of Sampling Set for Bandlimited Graph Signals}

\author{\IEEEauthorblockN{Xuan Xie, Junhao Yu, Hui Feng, Bo Hu}
\IEEEauthorblockA{\textit{Research Center of Smart Networks and Systems, School of Information Science and Technology, Fudan University} \\
Shanghai 200433, China \\
\{xxie15, jhyu17, hfeng, bohu\}@fudan.edu.cn}}

\maketitle

\begin{abstract}
The design of sampling set (DoS) for bandlimited graph signals (GS) has been extensively studied in recent years, but few of them exploit the benefits of the stochastic prior of GS.
In this work, we introduce the optimization framework for Bayesian DoS of bandlimited GS.
We also illustrate how the choice of different sampling sets affects the estimation error and how the prior knowledge influences the result of DoS compared with the non-Bayesian DoS by the aid of analyzing Gershgorin discs of error metric matrix.
Finally, based on our analysis, we propose a heuristic algorithm for DoS to avoid solving the optimization problem directly.
\end{abstract}


%
\IEEEpeerreviewmaketitle

\section{Introduction}
Sampling is a fundamental problem in graph signal processing (GSP) theory \cite{chen2015discrete,anis2016efficient,xie2017design,puy2018random,sakiyama2019eigendecomposition}. 
A signal can be recovered from partial samples only when we have some prior knowledge. 
There are usually three types of priors for signals: \textit{subspace prior, smoothness prior and stochastic prior} \cite{eldar2015sampling}. 
The subspace prior is the most widely used one in GSP which usually refers to bandlimitedness  or spectrally sparse. 
The smoothness prior cares about the energy distribution of graph signals (GS), for example, $p$-Dirichlet \cite{shuman2013emerging}  and total variation \cite{chen2015signal} in vertex domain and approximated bandlimited \cite{chen2016signal} in the spectral domain. 
The first and second-order statistics of a graph signal is required in the stochastic prior \cite{perraudin2017stationary,chamon2017greedy}. 
For example, a stochastic graph signal may follow joint zero-mean Gaussian distribution \cite{zhang2015graph,sakiyama2016efficient}. 

A qualified sampling set for reconstruction or estimation can be experimental design or non-experimental design.
Typical non-experimental design includes random sampling \cite{chen2016signal,chen2015signal} and topology based sampling \cite{puy2018random} which uses graph local coherence or leverage score as sampling score.
Experimental design usually tries to find sampling sets that minimize certain metric of the error covariance, such as A-optimal \cite{anis2016efficient,chamon2017greedy}, D-optimal \cite{anis2016efficient,chen2015discrete,xie2017design} and E-optimal \cite{anis2016efficient}. 
Other experimental designs aim to maximize the cut-off frequency \cite{anis2016efficient} or minimize the mean-square deviation \cite{di2018adaptive}. 

However, most of the experimental design are non-Bayesian, which means the stochastic prior of GS is not fully considered. 
Several related works which consider the stochastic prior only use it in the Bayesian estimator but not discuss the effects of it to the DoS.
A Bayesian estimator is used in \cite{chamon2017greedy} and the sampling set is chosen by minimizing the mean square error (MSE) of the estimation. 
But it focuses on proposing a greedy algorithm for Design of Sampling Set (DoS) but not discusses the connection between the DoS and the stochastic prior of GS.
Sensor selection problems \cite{shamaiah2010greedy,joshi2009sensor} which are formulated as Bayesian D-optimal, are similar to our work but they do not model the signals on sensors as GS.

In this work, we introduce the optimization framework for Bayesian DoS of bandlimited GS and show that most of the existing Bayesian DoS can be derived from our framework by choosing appropriate utility functions.
Inspired by \cite{bai2019reconstruction}, we analyze which vertices are preferred to be sampled for a known topology and priors and explain how stochastic prior makes the Bayesian DoS different from non-Bayesian DoS using Gershgorin circle theorem. 
However, the work of \cite{bai2019reconstruction} aims to design a sampling set that improve the condition number of the metric matrix of signal estimator, while we try to maximize the upper bound of eigenvalues of our metric matrix in the object function.
Finally, we propose a heuristic algorithm which does not need to solve an optimization problem for Bayesian DoS.   
Numerical simulation shows that our method has a smaller estimation error and is more robust against noise.

\section{Backgrounds}
Consider an $N$-vertex undirected connected graph $\mathcal{G}=(\mathcal{V},\mathcal{E},\mathbf{W})$, where $\mathcal{V}$ is the vertex set, $\mathcal{E}$ is the edge set and $\mathbf{W}$ is the weighted adjacency matrix. If there is an edge $e=(i,j)$ between vertices $i$ and $j$, then the entry $W_{i,j}$ represents the weight of the edge; otherwise $W_{i,j}=0$. A graph signal defined on $\mathcal{V}$ can be represented as a vector $\mathbf{f} \in \mathbb{C}^N$ with its element $f_i$ representing the signal value at the $i$th vertex.

The graph Laplacian is defined as $\mathbf{L}=\mathbf{D}-\mathbf{W}$, where the degree matrix is $\mathbf{D}=\text{diag}(\mathbf{W1})$. Since the Laplacian matrix is real symmetric, it has a complete eigenbasis and the spectral decomposition $\mathbf{L}=\mathbf{V}\mathbf{\Lambda}\mathbf{V}^T$,
where the eigenvectors  $\{\mathbf{v}_k\}_{1\leq k\leq N}$ of $\mathbf{L}$ form the columns of $\mathbf{V}$, and $\mathbf{\Lambda}$ is a diagonal matrix of eigenvalues $0=\lambda_1\leq \lambda_1\leq \cdots\leq \lambda_{N}$ of $\mathbf{L}$. 
The eigenvectors of the graph Laplacian are regarded as the Graph Fourier bases and the eigenvalues are regarded as frequencies \cite{sandryhaila2014discrete}. 
The expansion coefficients of a graph signal $\mathbf{f}$ corresponding to eigenvectors are defined as $\hat{\mathbf{f}}$, so that graph Fourier transform (GFT) can be expressed as
\begin{align}
\label{GFT}
\mathbf{f}=\mathbf{V}\hat{\mathbf{f}}.
\end{align}
\textbf{Subspace prior} \cite{chen2015discrete,anis2016efficient}: A graph signal can be represented by a linear combination of a subset of $\{\mathbf{v}_k\}$. Explicitly, if $\mathbf{f}$ is in the $\mathcal{K}$-subspace, where $\mathcal{K}\subset \mathcal{V}$ and $|\mathcal{K}| = K$, then it satisfies
\begin{equation}
\label{subspace}
\mathbf{f}=\mathbf{V}_{\mathcal{K}}\hat{\mathbf{f}}_{\mathcal{K}}.
\end{equation}
Note that bandlimitedness is a special case of subspace prior.\\
\textbf{Stochastic prior} \cite{perraudin2017stationary,chamon2017greedy,marques2017stationary,zhang2015graph}: $\hat{\mathbf{f}}_{\mathcal{K}}$ is known to be drawn from the following distribution
\begin{IEEEeqnarray}{Rl}
\label{stochastic}
p(\hat{\mathbf{f}}_{\mathcal{K}}) \propto \text{exp}(-(\hat{\mathbf{f}}_{\mathcal{K}} - \bm{\mu})^T\mathbf{\Sigma}^{-1}_{\hat{\mathbf{f}}_{\mathcal{K}}}(\hat{\mathbf{f}}_{\mathcal{K}} - \bm{\mu})),
\end{IEEEeqnarray}
where $\bm{\mu}$ is the mean of $\hat{\mathbf{f}}_{\mathcal{K}}$. Let $\mathbf{\Sigma}_{\hat{\mathbf{f}}_{\mathcal{K}}} = \text{diag}(\sigma^2_{\mathcal{K}_1},\cdots,\sigma^2_{\mathcal{K}_K})$, then each diagonal element represents the uncertainty of the corresponding mean value.
Without loss of generality, we assume that $\mathbf{\Sigma}_{\hat{\mathbf{f}}_{\mathcal{K}}}$ is full-rank. If $\bm{\mu}$ is constant over the vertex set, this prior is analogous to the generative model for a Gaussian wide-sense stationary random process on graphs \cite{perraudin2017stationary}, but we do not restrict $\bm{\mu}$ to be constant. 
For example, in a sensor network measuring the temperature of a large area, the mean temperature may vary across different locations.
In this paper, we assume that the stochastic prior is known. How to estimate the prior is an independent problem which is studied in \cite{perraudin2017stationary,marques2017stationary}.

Suppose that we sample $M$ measurements from the graph signal $\mathbf{f}$ to produce a sampled signal $\mathbf{f}_{\mathcal{S}}\in \mathbb{C}^M$, usually $M < N$, where $\mathcal{S}=(\mathcal{S}_1,\cdots,\mathcal{S}_M)$ denotes the sequence of sampled indices, and $\mathcal{S}_i\in \{1,2,\cdots,N \}$. The sampling operator $\mathbf{\Psi}:\mathbb{C}^N\mapsto \mathbb{C}^M$ is defined as
\begin{equation}
\label{sampling operator}
\mathbf{\Psi}_{i,j}=
        \left\{\begin{matrix}
        1, & j=\mathcal{S}_i;\\
        0, & \text{otherwise.}
        \end{matrix}\right.
\end{equation}
The observation model is $\mathbf{y}_{\mathcal{S}}=\mathbf{\Psi}\mathbf{y} = \mathbf{\Psi}(\mathbf{V}_\mathcal{K}\hat{\mathbf{f}}_\mathcal{K}+\mathbf{w})$, where $\mathbf{w}$ is the \textit{i.i.d} noise with zero mean and covariance matrix $\mathbf{\Sigma}_{\mathbf{w}} = \sigma_{\mathbf{w}}^2\mathbf{I}$.
\section{Framework}
In this paper, the goal is to design a sampling set that can estimate $\hat{\mathbf{f}}_{\mathcal{K}}$ by samples on them with the least estimation error. 
If $m_i$ observations are taken at the $i$-th vertex, then we define the design as ${\eta}_i = m_i/M$, which is a proportion of vertices being sampled. 
Inspired by \cite{chaloner1995bayesian}, a general utility function that describes the goal of DoS can be defined as $U(\hat{\mathbf{f}}_{\mathcal{K}},\bm{\eta},\mathbf{y}_{\mathcal{S}})$. 
For any design $\bm{\eta} = \{ \eta_1,\eta_2,\cdots,\eta_N\}$, the expected utility is given by averaging over what is unknown
\begin{equation}
\label{Utility}
U(\bm{\eta}) \triangleq \iint U(\hat{\mathbf{f}}_{\mathcal{K}},\bm{\eta},\mathbf{y}_{\mathcal{S}})p(\hat{\mathbf{f}}_{\mathcal{K}}|\mathbf{y}_{\mathcal{S}},\bm{\eta})p(\mathbf{y}_{\mathcal{S}}|\bm{\eta})d\hat{\mathbf{f}}_{\mathcal{K}} d\mathbf{y}_{\mathcal{S}},
\end{equation}
where $p(\cdot)$ denotes probability density function. 
The Bayesian solution is provided by the design that maximizes (\ref{Utility}).
Select the utility function that describes the information gain in $\hat{\mathbf{f}}_{\mathcal{K}}$ after sampling, then the expected utility can be given as 
\begin{align}
U_1(\bm{\eta}) \triangleq \iint  \log\frac{p(\hat{\mathbf{f}}_{\mathcal{K}}|\mathbf{y}_{\mathcal{S}},\bm{\eta})}{p(\hat{\mathbf{f}}_{\mathcal{K}})}p(\mathbf{y}_{\mathcal{S}},\hat{\mathbf{f}}_{\mathcal{K}}|\bm{\eta})d\hat{\mathbf{f}}_{\mathcal{K}} d\mathbf{y}_{\mathcal{S}},
\end{align}
which is also known as Mutual Information (MI).

The observations follow the Gaussian distribution $p(\mathbf{y}_{\mathcal{S}}|\hat{\mathbf{f}}_{\mathcal{K}};\mathbf{\Sigma}_{\mathbf{w}})= \mathcal{N}(\mathbf{\Psi}\mathbf{V}_\mathcal{K}\hat{\mathbf{f}}_\mathcal{K},\mathbf{\Psi}\mathbf{\Sigma}_{\mathbf{w}} \mathbf{\Psi}^T)$. 
If the prior distribution for $\hat{\mathbf{f}}_{K}$ is also Gaussian as (\ref{stochastic}), the posterior distribution for $\hat{\mathbf{f}}_\mathcal{K}$ is $p(\hat{\mathbf{f}}_\mathcal{K}|\mathbf{y}_{\mathcal{S}},\bm{\eta})= \mathcal{N}(\hat{\mathbf{f}}_\mathcal{K}^*,\mathbf{\Sigma}_{\text{B}}^*) $, where $\hat{\mathbf{f}}_\mathcal{K}^* = \mathbf{\Sigma}_{\text{B}}^*(\mathbf{V}_\mathcal{K}^T\mathbf{\Psi}^T\mathbf{\Sigma}_{\mathbf{w}}^{-1}\mathbf{y}_{\mathcal{S}} + \mathbf{\Sigma}_{\hat{\mathbf{f}}_{\mathcal{K}}}^{-1}\bm{\mu})$ is the minimum mean square error (MMSE) estimator of $\hat{\mathbf{f}}_{\mathcal{K}}$ and $\mathbf{\Sigma}_{\text{B}}^* = (\sigma_{\mathbf{w}}^{-2}\mathbf{V}_\mathcal{K}^T\mathbf{\Psi}^T\mathbf{\Psi}\mathbf{V}_\mathcal{K} + \mathbf{\Sigma}_{\hat{\mathbf{f}}_{\mathcal{K}}}^{-1})^{-1}$. 
We can find that $\mathbf{\Psi}^T\mathbf{\Psi}$ is a diagonal matrix and its $i$-th diagonal element equals $M\eta_i$ which represents how many times the $i$-th vertex is sampled.

Let $\mathbf{u}_1^T,\cdots,\mathbf{u}_N^T$ be the rows of $\mathbf{V}_{\mathcal{K}}$, then 
$\mathbf{\Sigma}_{\text{B}}^*$ is a function of the design $\bm{\eta}$ as
\begin{align}
\mathbf{\Sigma}^*_{\text{B}}(\bm{\eta}) \triangleq (\sigma_{\mathbf{w}}^{-2}M\sum_{i = 1}^N{\eta}_i\mathbf{u}_i\mathbf{u}_i^T+ \mathbf{\Sigma}_{\hat{\mathbf{f}}_{\mathcal{K}}}^{-1})^{-1}.
\end{align}
Since the prior distribution does not depends on the design $\bm{\eta}$, the design maximizes the expected utility is the one that maximizes 
\begin{align}
\label{expected utility U1}
U_1(\bm{\eta}) & \triangleq \iint \log p(\hat{\mathbf{f}}_\mathcal{K}|\mathbf{y}_{\mathcal{S}},\bm{\eta})p(\mathbf{y}_{\mathcal{S}},\hat{\mathbf{f}}_\mathcal{K}|\bm{\eta})d\hat{\mathbf{f}}_\mathcal{K} d\mathbf{y}_{\mathcal{S}}\nonumber\\
& = -\frac{K}{2}\log (2\pi) - \frac{K}{2} + \frac{1}{2}\log\det(\mathbf{\Sigma}_{\text{B}}^*(\bm{\eta}))^{-1}.
\end{align}
Therefore the Bayesian D-optimal for sensor selection in \cite{shamaiah2010greedy} can be derived from our framework. 

We can also select a utility function that describes the negative quadratic loss of the estimation. Assume that $\hat{\mathbf{f}}_\mathcal{K}$ is estimated by the MMSE estimator, the excepted utility is 
\begin{align}
\label{expected utility U2}
U_2(\bm{\eta}) &\triangleq -\iint (\hat{\mathbf{f}}_\mathcal{K} - \hat{\mathbf{f}}_\mathcal{K}^*)^T\mathbf{A}(\hat{\mathbf{f}}_\mathcal{K} - \hat{\mathbf{f}}_\mathcal{K}^*)p(\mathbf{y}_{\mathcal{S}},\hat{\mathbf{f}}_\mathcal{K}|\bm{\eta})d\hat{\mathbf{f}}_\mathcal{K} d\mathbf{y}_{\mathcal{S}}\nonumber\\
& = -\text{tr}\left(\mathbf{A}\mathbf{\Sigma}_{\text{B}}^*(\bm{\eta})\right),
\end{align}
where $\mathbf{A}$ is a symmetric positive semi-definite matrix. The object function in \cite{chamon2017greedy} can be derived from our framework by setting $\mathbf{A} = \mathbf{I}$.

The optimization problem of (\ref{expected utility U1}) and (\ref{expected utility U2}) can be expressed as follow
\vspace{-0.3cm}
\begin{align}
\label{optimization}
\max_{\bm{\eta}} \quad &U_1(\bm{\eta})\quad \text{or} \quad U_2(\bm{\eta})\nonumber\\
\text{s.t.} \quad & 0 \leq \eta_i \leq 1,\,\sum_{i = 1}^N{\eta}_i = 1,\nonumber\\
& M{\eta}_i\in \mathbf{Z}.
\end{align}

We can find that both (\ref{expected utility U1}) and (\ref{expected utility U2}) are related to the matrix $\mathbf{\Sigma}^*_{\text{B}}(\bm{\eta})$ which obviously depends on the sample size $M$. 
Recall that object functions in non-Bayesian experimental design \cite{chen2015discrete,anis2016efficient,xie2017design} are all related to the covariance matrix of Least Square (LS) estimator $\mathbf{\Sigma}_{\text{nB}}^* \triangleq (\sigma_{\mathbf{w}}^{-2}\mathbf{V}_\mathcal{K}^T\mathbf{\Psi}^T\mathbf{\Psi}\mathbf{V}_\mathcal{K})^{-1}$,
 which is also a function of design $\bm{\eta}$ since $\mathbf{\Sigma}_{\text{nB}}^*(\bm{\eta}) = (M\sigma_{\mathbf{w}}^{-2}\sum_{i=1}^N\eta_i\mathbf{u}_i\mathbf{u}_i^T)^{-1}$.  
If $M$ is large, the object function in our method will not be sensitive to the prior distribution $\mathbf{\Sigma}_{\hat{\mathbf{f}}_{\mathcal{K}}}^{-1}$ and will be approximately equal to the non-Bayesian one. 
In contrast, if $M$ is small, the prior distribution will have more effect on the DoS. 
So, if the stochastic prior to GS is informative, our approach will need fewer samples to achieve the same goal compared to the non-Bayesian approach. 
But if the prior is noninformative, there will be no advantage using the Bayesian DoS for sampling. 

There is another advantage of using Bayesian DoS. 
In non-Bayesian DoS, the sampling set may lead to a singular $\mathbf{\Sigma}^*_{\text{nB}}(\bm{\eta})$. 
To avoid this, some adaptation is required. A possible way is to increase the sample size to ensure the nonsingularity of $\mathbf{\Sigma}^*_{\text{nB}}(\bm{\eta})$ of a given probability \cite{xie2017design}. 
However in Bayesian DoS, the matrix $\mathbf{\Sigma}^*_{\text{B}}(\bm{\eta})$ is always nonsingular as long as we choose a nonsingular $\mathbf{\Sigma}_{\hat{\mathbf{f}}_{\mathcal{K}}}$.
\section{Algorithm}
\vspace{-0.2cm}
\subsection{Convex relaxation}
The optimization problem (\ref{optimization}) is an intractable combinatorial problem, but it can be converted to a convex optimization problem by relaxing the constraint condition $ M{\eta}_i\in \mathbf{Z}$ \cite{boyd2004convex}, 
\begin{align}
\label{relaxed optimization}
\max_{\bm{\eta}} \quad &U_1(\bm{\eta})\quad \text{or} \quad U_2(\bm{\eta})\nonumber\\
\text{s.t.} \quad & 0 \leq \eta_i \leq 1,\,\sum_{i = 1}^N{\eta}_i = 1.
\end{align}
Problem (\ref{relaxed optimization}) can be solved by any optimization tool like interior-point methods and the solution is denoted by $\bm{\eta}^*$. The optimal value of (\ref{relaxed optimization}) provides a lower bound on the optimal value of (\ref{optimization}). 
\vspace{-0.1cm}
\subsection{Difference between non-Bayesian and Bayesian DoS}
\label{DoSBnB}
In this section, based on the problem (\ref{relaxed optimization}), we will take utility function $U_1$ as an example to make a qualitative analysis about the difference between non-Bayesian and Bayesian DoS and explain the reason.
Instead of solving (\ref{relaxed optimization}) directly with convex optimization tools, we intend to analyze how $\bm{\eta}$ changes the bound of eigenvalues of $(\mathbf{\Sigma}_{\text{nB}}^*(\bm{\eta}))^{-1}$ and $(\mathbf{\Sigma}_{\text{B}}^*(\bm{\eta}))^{-1}$, and find a heuristic method to decide $\bm{\eta}^*$ with low complexity.
Let $\mathbf{A} = \mathbf{C} + \mathbf{R}$ be a $K\times K$ matrix, where $\mathbf{C}$ is a diagonal matrix and $\mathbf{R}$ is a matrix with all zeros on the diagonal. Then, $\log\det(\mathbf{A}) = \sum_{i=1}^K\log \lambda_i$, where $\lambda_1,\cdots,\lambda_K$ are the eigenvalues of $\mathbf{A}$. 
From Gershgorin circle theorem, each eigenvalue of $\mathbf{A}$ lies within at least one of the Gershgorin discs $GD(c_{ii},R_i)$ with disc center $c_{ii}$ and radius $R_i$, where $R_i = \sum_{j\neq i}|R_{ij}|$.

Non-Bayesian DoS in \cite{xie2017design} seeks for an optimal sampling proportion $\bm{\eta}^*$ that maximizes 
\begin{IEEEeqnarray}{Rl}
\log\det(\mathbf{\Sigma}_{\text{nB}}^*(\bm{\eta}))^{-1} & = \log\det(\sigma_{\mathbf{w}}^{-2}M\mathbf{V}_{\mathcal{K}}^T\text{diag}(\bm{\eta})\mathbf{V}_{\mathcal{K}})\IEEEnonumber\\
 & = \log\det(\sigma_{\mathbf{w}}^{-2}) + \sum_{i=1}^K\log(\lambda^{\text{nB}}_i),
\label{SigmanB}
\end{IEEEeqnarray}
where $\bm{\lambda}^{\text{nB}} = \{\lambda^{\text{nB}}_1,\cdots,\lambda^{\text{nB}}_K\}$ denote the eigenvalues of $(M\mathbf{V}_{\mathcal{K}}^T\text{diag}(\bm{\eta})\mathbf{V}_{\mathcal{K}})$, which is the main part of $\mathbf{\Sigma}_{\text{nB}}^*(\bm{\eta})$.
And $\bm{\lambda}^{\text{nB}}$ corresponding to $\bm{\eta}^*$ lies within the Gershgorin discs of $(M\mathbf{V}_{\mathcal{K}}^T\text{diag}(\bm{\eta}^*)\mathbf{V}_{\mathcal{K}})$.
However, each center or radius of Gershgorin discs of $(M\mathbf{V}_{\mathcal{K}}^T\text{diag}(\bm{\eta}^*)\mathbf{V}_{\mathcal{K}})$ is decided by the sampling proportion of all vertices together, which makes it hard to analysis.
Instead, we analyze the bound of $\bm{\lambda}^{\text{nB}}$ through Gershgorin discs of
\begin{IEEEeqnarray}{Rl}
\mathbf{G}&_{\text{nB}}(\bm{\eta}^*) \triangleq M\text{diag}(\bm{\eta}^*)^{\frac{1}{2}}\mathbf{V}_{\mathcal{K}}\mathbf{V}_{\mathcal{K}}^T\text{diag}(\bm{\eta}^*)^{\frac{1}{2}} \IEEEnonumber\\
&=M\text{diag}(\bm{\eta}^*)^{\frac{1}{2}}
\begin{bmatrix}\mathbf{u}_1^T\mathbf{u}_1 & \cdots  & \mathbf{u}_1^T\mathbf{u}_N\\ 
\vdots  & \ddots   & \vdots \\ 
\mathbf{u}_1^T\mathbf{u}_N & \cdots  & \mathbf{u}_N^T\mathbf{u}_N
\end{bmatrix}
\text{diag}(\bm{\eta}^*)^{\frac{1}{2}}
,\IEEEnonumber
\end{IEEEeqnarray}
since its nonzero eigenvalues are the same as $\bm{\lambda}^{\text{nB}}$. 
By doing so, we simplify the mathematics form of diagonal elements and make it more clear about how $\bm{\eta}$ changes the disc centers.
Obviously, the Gershgorin discs of $\mathbf{G}_{\text{nB}}(\bm{\eta}^*)$ can be obtained by changing the discs of $(\mathbf{V}_{\mathcal{K}}\mathbf{V}_{\mathcal{K}}^T)$ which is related to the error metric matrix of uniform sampling, from $GD(\mathbf{u}_i^T\mathbf{u}_i, \sum_{j\neq i}|\mathbf{u}_i^T\mathbf{u}_j|)$ to $GD(M{\eta}_i^*\mathbf{u}_i^T\mathbf{u}_i,M\sum_{j\neq i}|\sqrt{\eta_i^*\eta_j^*}\mathbf{u}_i^T\mathbf{u}_j|)$ by $\bm{\eta}^*$.

In Bayesian DoS, we seeks for the $\bm{\eta}^*$ that maximizes 
\begin{IEEEeqnarray}{Rl}
\label{GD_B}
&\log\det(\mathbf{\Sigma}_{\text{B}}^*(\bm{\eta}))^{-1} \IEEEnonumber\\
&=\log\det(\sigma_\mathbf{w}^{-2}\mathbf{\Sigma}_{\hat{\mathbf{f}}_{\mathcal{K}}}^{-\frac{1}{2}}(M\mathbf{\Sigma}_{\hat{\mathbf{f}}_{\mathcal{K}}}^{\frac{1}{2}}\mathbf{V}_{\mathcal{K}}^T\text{diag}(\bm{\eta})\mathbf{V}_{\mathcal{K}}\mathbf{\Sigma}_{\hat{\mathbf{f}}_{\mathcal{K}}}^{\frac{1}{2}} + \sigma_\mathbf{w}^2\mathbf{I})\mathbf{\Sigma}_{\hat{\mathbf{f}}_{\mathcal{K}}}^{-\frac{1}{2}})\IEEEnonumber\\
& = \log\det(\sigma_\mathbf{w}^{-2}\mathbf{\Sigma}_{\hat{\mathbf{f}}_{\mathcal{K}}}^{-1}) + \sum_{i=1}^K\log(\lambda_i^{\text{B}}),
\end{IEEEeqnarray}
where $\{\lambda^{\text{B}}_1,\cdots,\lambda^{\text{B}}_K\}$ are the eigenvalues of $(M\mathbf{\Sigma}_{\hat{\mathbf{f}}_{\mathcal{K}}}^{\frac{1}{2}}\mathbf{V}_{\mathcal{K}}^T\text{diag}(\bm{\eta})\mathbf{V}_{\mathcal{K}}\mathbf{\Sigma}_{\hat{\mathbf{f}}_{\mathcal{K}}}^{\frac{1}{2}} + \sigma_\mathbf{w}^2\mathbf{I})$.
In order to analyze the bound of $\bm{\lambda}^{\text{B}}$ corresponding to $\bm{\eta}^*$, we define 
\begin{IEEEeqnarray}{Rl}
\label{GB}
\mathbf{G}_{\text{B}}(\bm{\eta}) \triangleq M\text{diag}(\bm{\eta})^{\frac{1}{2}} \mathbf{V}_{\mathcal{K}}\mathbf{\Sigma}_{\hat{\mathbf{f}}_{\mathcal{K}}}\mathbf{V}_{\mathcal{K}}^T\text{diag}(\bm{\eta})^{\frac{1}{2}}+ \sigma_\mathbf{w}^2\mathbf{I},
\end{IEEEeqnarray}
whose $K$ largest eigenvalues are the same as $\bm{\lambda}^{\text{B}}$ and the remaining eigenvalues are constants irrelevant to $\bm{\eta}$.
Since the first term in (\ref{GD_B}) is not related to the design, the optimal $\bm{\eta}^*$ that maximizes (\ref{GD_B}) is the one that maximizes $\log\det(\mathbf{G}_{\text{B}}(\bm{\eta}))$. 
Recall (\ref{GB}) and the analysis for non-Bayesian DoS, $\bm{\lambda}^{\text{B}}$ lies within Gershgorin discs of $\mathbf{G}_{\text{B}}(\bm{\eta}^*)$ which can be obtained by changing the discs of $(\mathbf{V}_{\mathcal{K}}\mathbf{\Sigma}_{\hat{\mathbf{f}}_{\mathcal{K}}}\mathbf{V}_{\mathcal{K}}^T)$ by $\bm{\eta}^*$ and then translating all the discs by $\sigma_\mathbf{w}^2$. 

\subsection{Bayesian graph coherence}
For some large scale GS, it is expensive to solve (\ref{relaxed optimization}) directly. So, heuristic methods are preferred to decide $\bm{\eta}^*$. 
According to the analysis in Section \ref{DoSBnB}, for non-Bayesian DoS, we can design an $\bm{\eta}^*$ to change the discs of $(\mathbf{V}_{\mathcal{K}}\mathbf{V}_{\mathcal{K}}^T)$, which raises both the lower bound and the upper bound of nonzero $\bm{\lambda}^{\text{nB}}$. An alternative approach is to give a large sampling proportion ${\eta}^*_i$ to the $i$th vertex if the corresponding center $\mathbf{u}_i^T\mathbf{u}_i$ is large. 
This is analogous to the sampling strategy in \cite{puy2018random}.  
Let $\bm{\delta}_i$ be a graph signal with value 1 at vertex $i$ and 0 everywhere else.
Then, the center $\mathbf{u}_i^T\mathbf{u}_i = \Vert \mathbf{V}_{\mathcal{K}}^T\bm{\delta}_i\Vert_2^2$ is defined as the local graph coherence \cite{puy2018random} at vertex $i$, which characterises how much the energy of $\bm{\delta}_i$ is concentrated in the $\mathcal{K}$-subspace.

Denoting the rows of $(\mathbf{V}_\mathcal{K}\mathbf{\Sigma}_{\hat{\mathbf{f}}_{\mathcal{K}}}^{1/2})$ by $\tilde{\mathbf{u}}_1^T,\cdots,\tilde{\mathbf{u}}_N^T$, we define the \textit{Bayesian graph coherence} for vertex $i$ as $\tilde{\mathbf{u}}_i^T\tilde{\mathbf{u}}_i = \Vert \mathbf{\Sigma}_{\hat{\mathbf{f}}_{\mathcal{K}}}^{1/2}\mathbf{V}_\mathcal{K}^T \bm{\delta}_i\Vert_2^2$, which is the weighted energy of $\bm{\delta}_i$ concentrated in the $\mathcal{K}$-subspace.  
To introduce a heuristic method for Bayesian DoS, we give a large sampling proportion ${\eta}^*_i$ to the vertex $i$ if the center $\tilde{\mathbf{u}}_i^T\tilde{\mathbf{u}}_i$ is large to raise both the lower and the upper bound of $\bm{\lambda}^{\text{B}}$. 
Since $\Vert \mathbf{V}_\mathcal{K}\mathbf{\Sigma}_{\hat{\mathbf{f}}_{\mathcal{K}}}^{1/2}\Vert_2^2 = \sum_{i=1}^K \sigma^2_{\mathcal{K}_i}$, the sampling proportion in our heuristic method is given as 
\begin{IEEEeqnarray}{Rl}
\label{BayeCoherence}
{\eta}_i^* = \frac{\tilde{\mathbf{u}}_i^T\tilde{\mathbf{u}}_i}{\sum_{i=1}^K \sigma^2_{\mathcal{K}_i}}
\end{IEEEeqnarray}
to ensure that $\sum_{i=1}^K {\eta}_i^* = 1$.
The sampling proportion in (\ref{BayeCoherence}) indicates that a vertex that has a large signal amplitude in the spectral components of which the GFT coefficients have a large variance in prior knowledge is preferred to be sampled.
In other words, we sample the vertices to reduce the uncertainty of $\hat{\mathbf{f}}_{\mathcal{K}}$.
Different from the greedy algorithm in \cite{shamaiah2010greedy,chamon2017greedy} and the convex relaxation in \cite{joshi2009sensor} which all focus on solving the optimization problem numerically, our heuristic algorithm gives an explanation about why a vertex should be sampled for a given topology, which is relevant to $\mathbf{V}$, and prior knowledge.

However, to know how many times each vertex should be sampled, each entry of $\bm{\eta}^*$ need to be quantified to an integer multiple of $\frac{1}{M}$. 
A probabilistic quantization method proposed in our previous work \cite{xie2017design} can be applied.
\section{Experiments}
\begin{figure}[!t]
\subfigure[$\mathbf{v}_{10}$]
{
\begin{minipage}[b]{.28\linewidth}
  \centering
  \includegraphics[width=\linewidth]{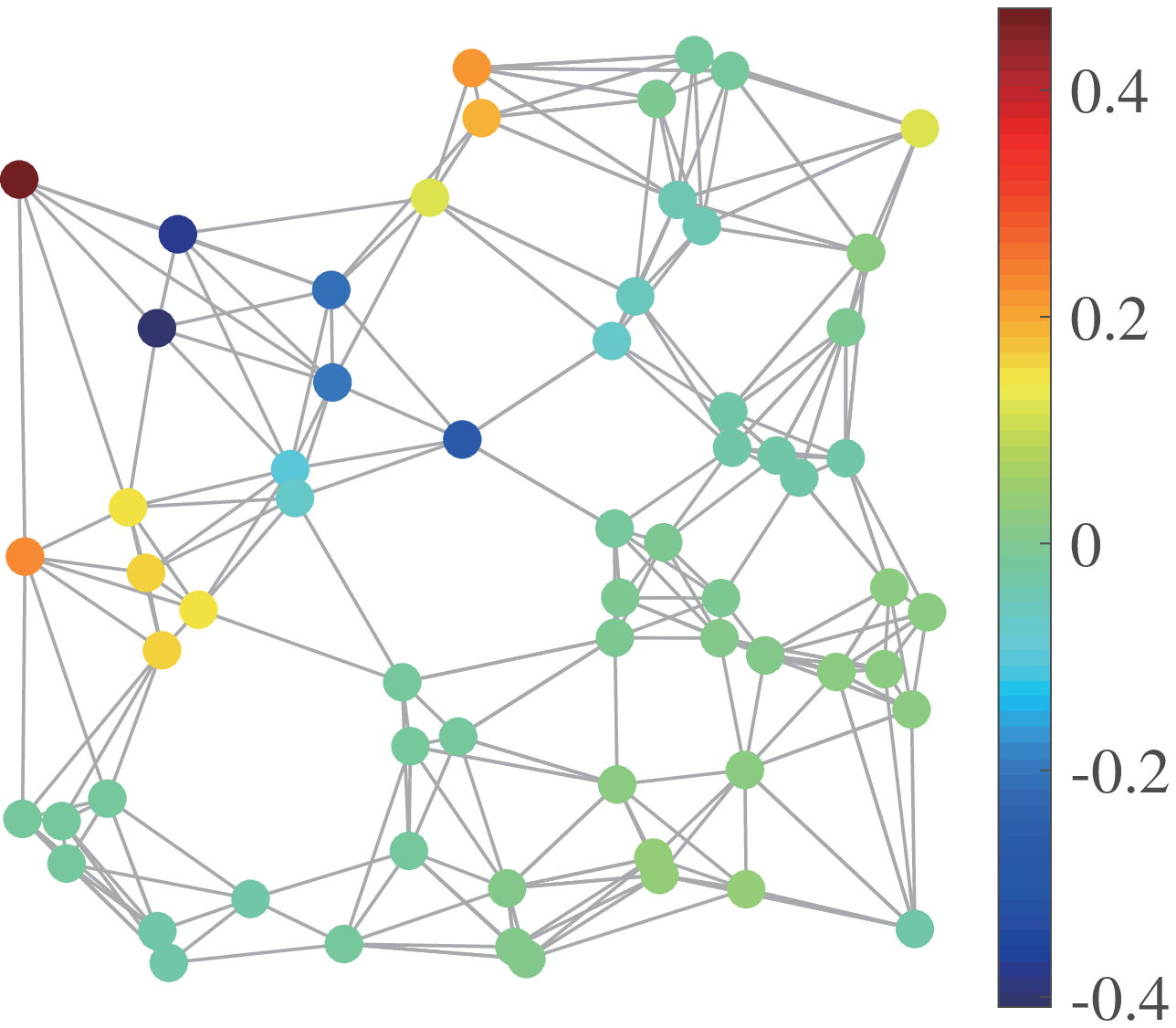}
\end{minipage}
}
\hfill
\subfigure[$\mathbf{v}_{20}$]
{
\begin{minipage}[b]{0.28\linewidth}
  \centering
  \includegraphics[width=\linewidth]{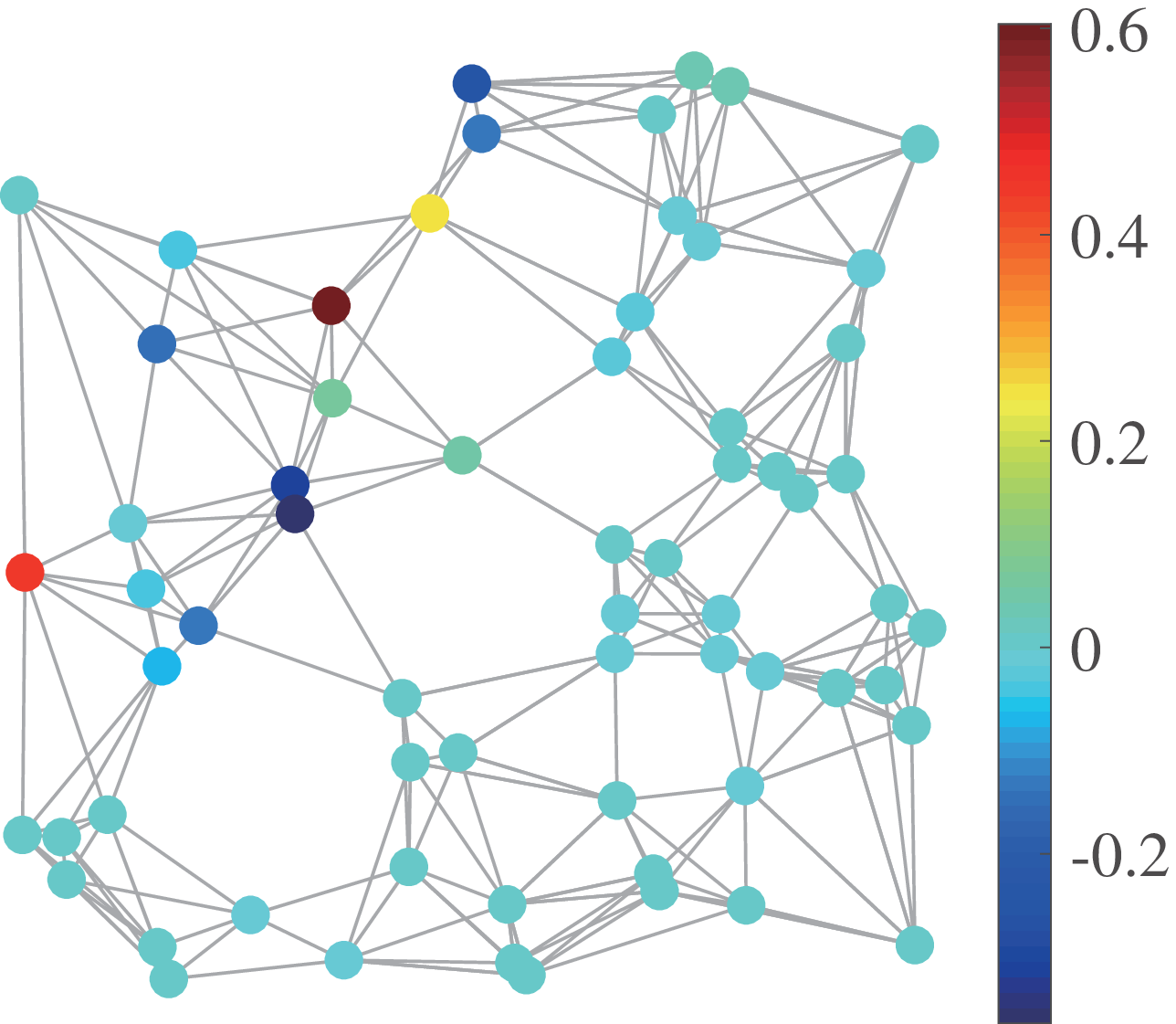}
\end{minipage}
}
\hfill
\subfigure[$\mathbf{v}_{30}$]
{
\begin{minipage}[b]{.28\linewidth}
  \centering
  \includegraphics[width=\linewidth]{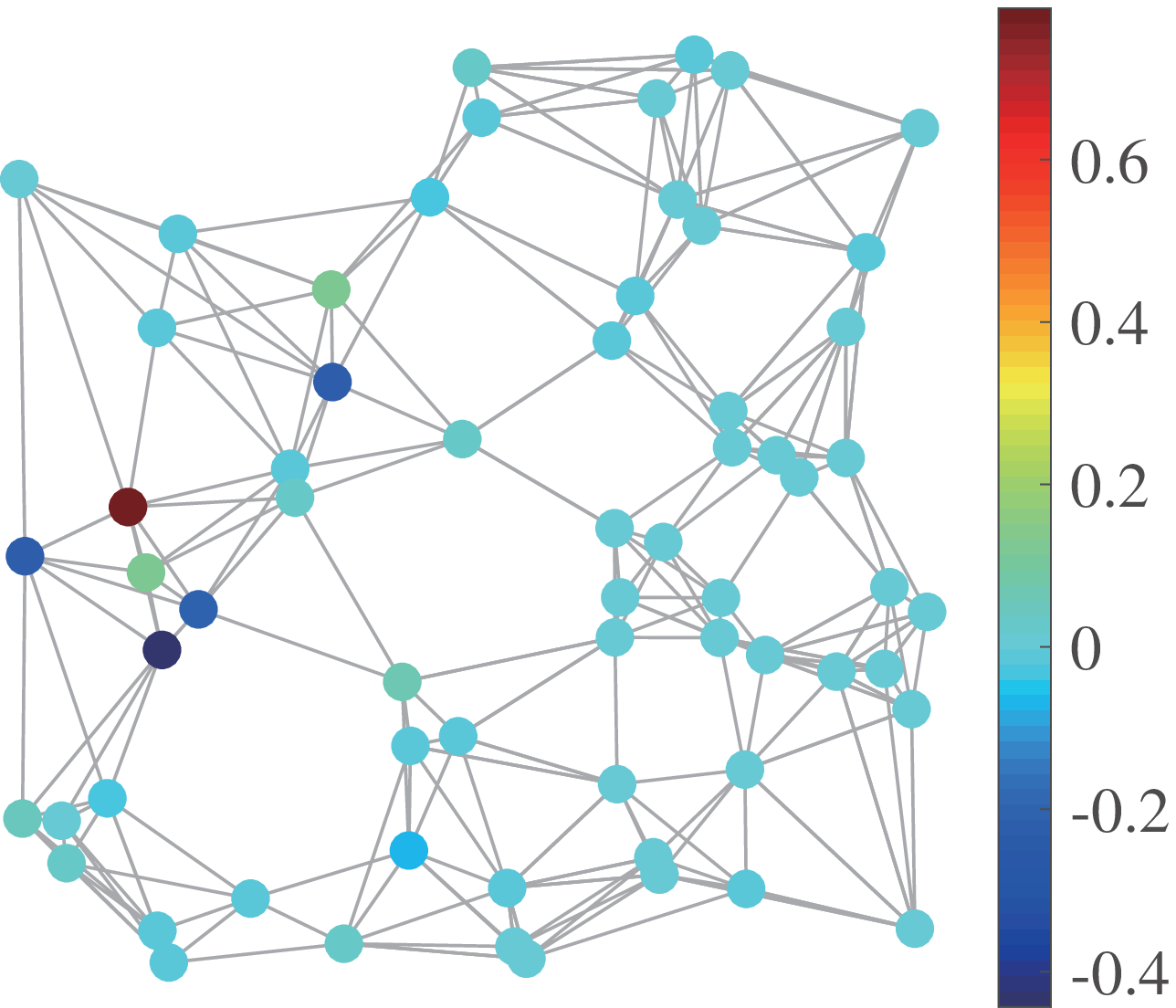}
\end{minipage}
}

~

\centering
\hfill
\subfigure[Result for non-Bayesian DoS]
{
\begin{minipage}[b]{.45\linewidth}
  \centering
  \includegraphics[scale = 0.2]{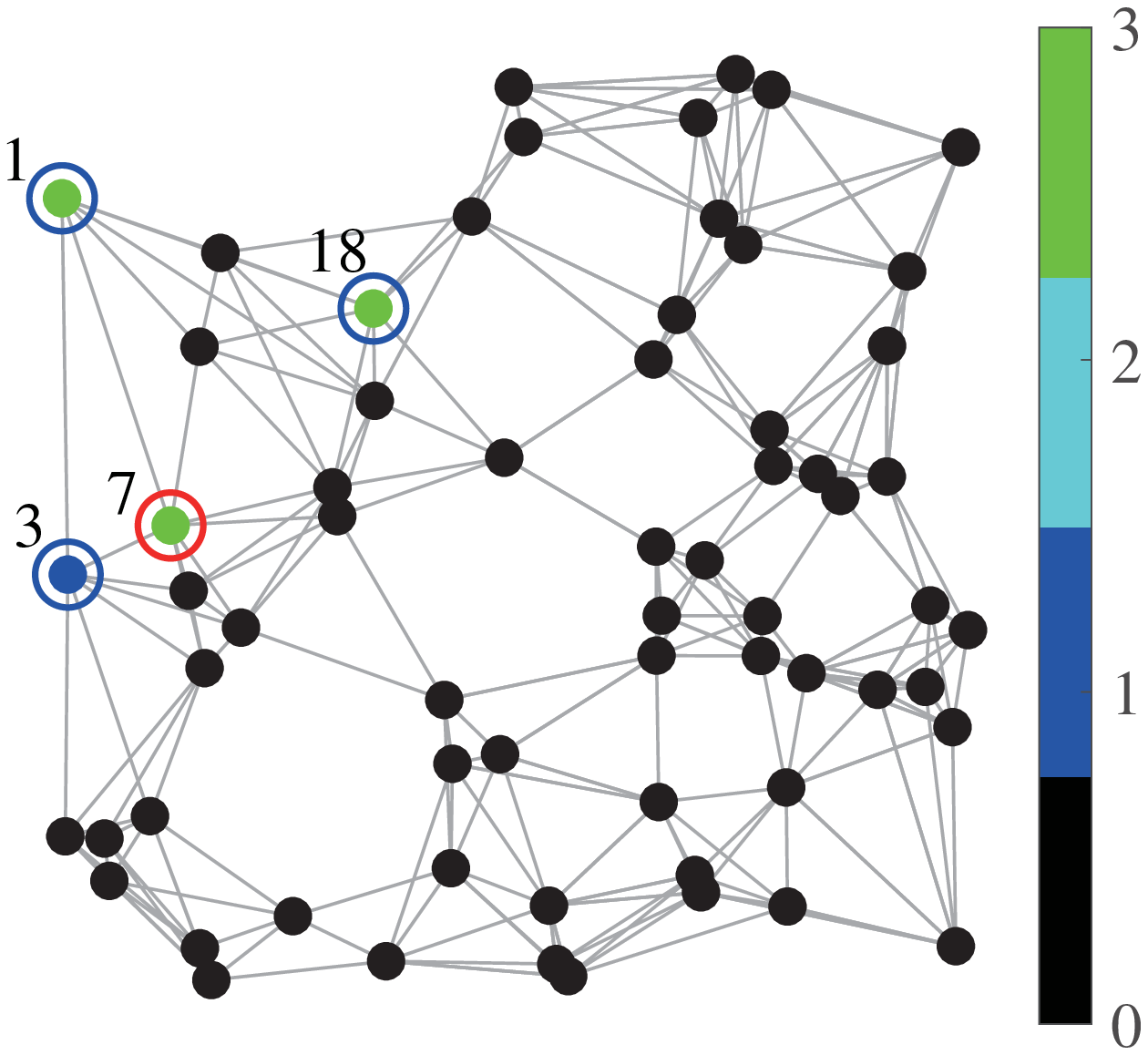}
\end{minipage}
}
\hfill
\subfigure[Result for Bayesian DoS]
{
\begin{minipage}[b]{0.45\linewidth}
  \centering
  \includegraphics[scale = 0.2]{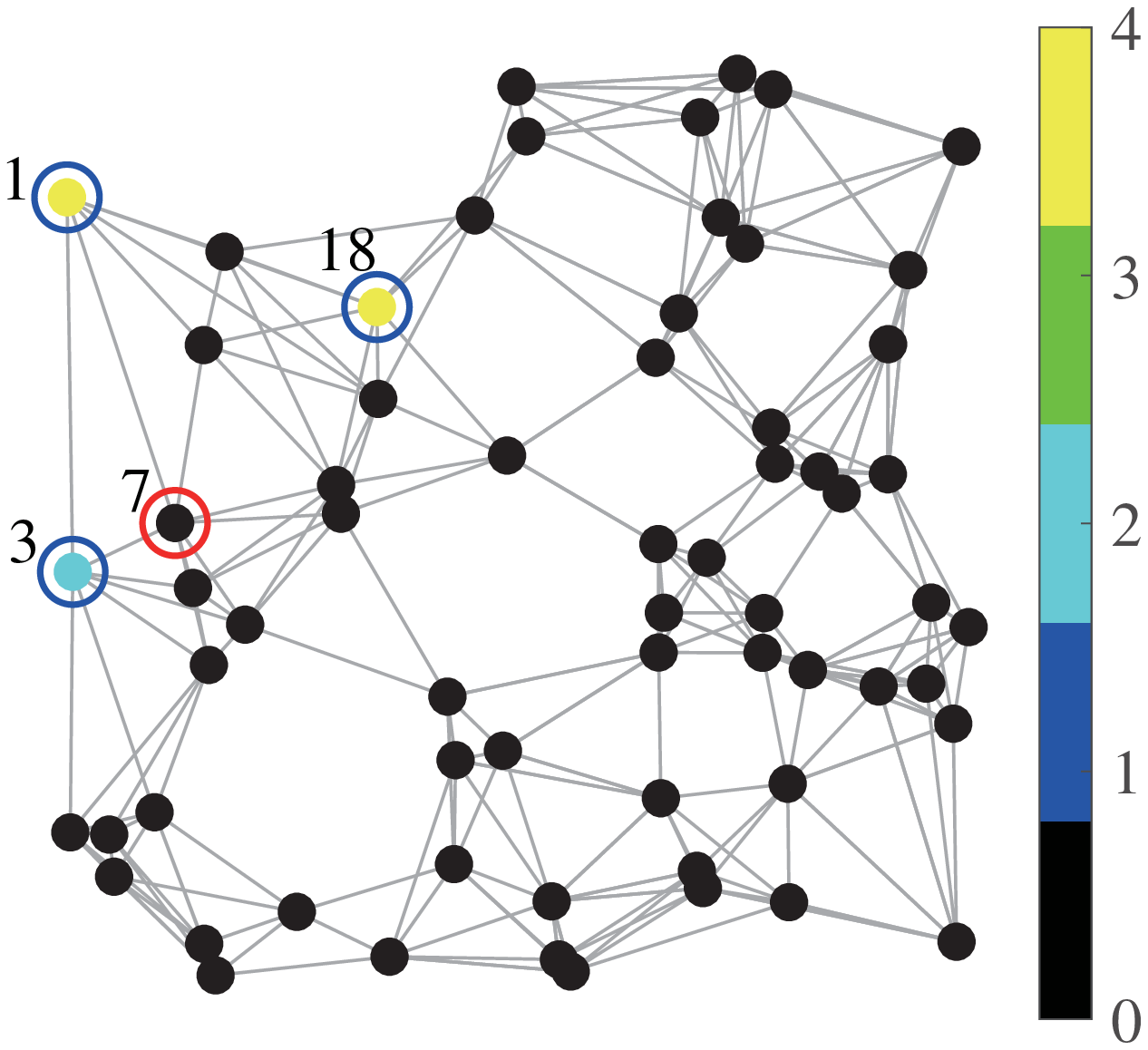}
  \end{minipage}
}
\hfill
\caption{Results for DoS. In (a),(b) and (c), the color bar represents the signal value on each vertex and in (d) and (e) it represents the sampling size on each vertex.}
\label{difference}
\end{figure}

First, we illustrate the difference of DoS between non-Bayesian and Bayesian on a random geometric graph with 64 vertices placed randomly in the unit square, and edges are placed between any vertices within 0.6. 
The edge weights are assigned via a Gaussian kernel. The graph signal lies in the column space of $\mathbf{V}_{\mathcal{K}}$ with $\mathcal{K} = \{10,20,30\}$ which are shown in Fig. \ref{difference}(a),(b),(c). 
The sampling budget $M=10$ and the samples are noisy with additive \textit{i.i.d.} Gaussian noise with $\sigma_{\mathbf{w}}^2 = 0.5$. 
The mean of $\hat{\mathbf{f}}_{\mathcal{K}}$ is $\mathbf{\mu} = \mathbf{1}_{3\times 1}$ and the covariance matrix is $\mathbf{\Sigma}_{\hat{\mathbf{f}}_{\mathcal{K}}} = \text{diag}(1,0.5,0.1)$. 
The result for DoS of \cite{xie2017design} is shown in Fig. \ref{difference}(d) and the result for our Bayesian DoS is shown in Fig. \ref{difference}(e).  
We can find that the vertex in the red circle is sampled for 3 times in Fig. \ref{difference}(d) but not sampled in Fig. \ref{difference}(e) since the signal amplitude of it in $\mathbf{v}_{10}$ and $\mathbf{v}_{20}$ is small, while the vertices in the blue circle are all sampled for one more time.

\begin{figure}[!t]
  \centering
\subfigure[Centers and radiuses of Gershgorin discs for $\mathbf{V}_{\mathcal{K}}\mathbf{V}_{\mathcal{K}}^T$ (blue) and $\mathbf{G}_{\text{nB}}(\bm{\eta}^*)$ (red).]
{
\begin{minipage}[b]{.45\linewidth}
  \centering
  \includegraphics[scale = 0.48]{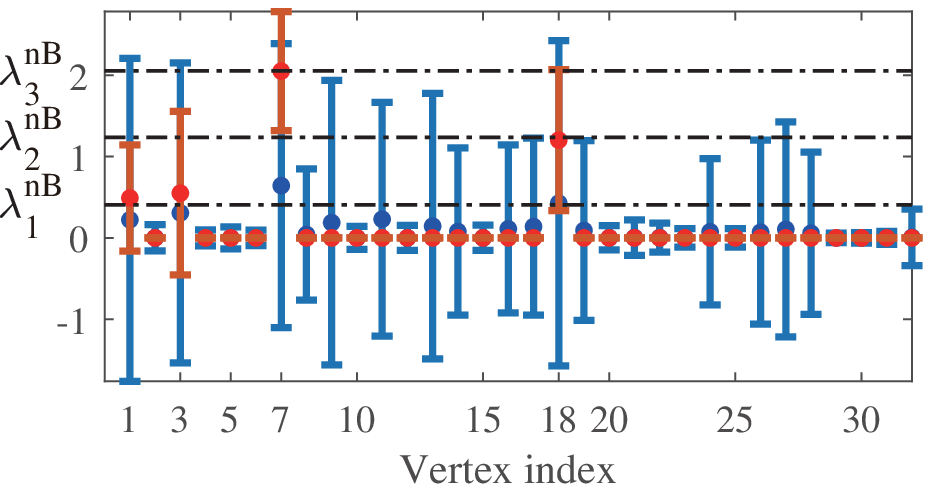}
\end{minipage}
}
\hfill
\subfigure[Centers and radiuses of Gershgorin discs for $\mathbf{V}_{\mathcal{K}}\mathbf{\Sigma}_{\hat{\mathbf{f}}_{\mathcal{K}}}\mathbf{V}_{\mathcal{K}}^T$ (blue) and $\mathbf{G}_{\text{B}}(\bm{\eta}^*)$ (red).]
{
\begin{minipage}[b]{0.45\linewidth}
  \centering
  \includegraphics[scale = 0.48]{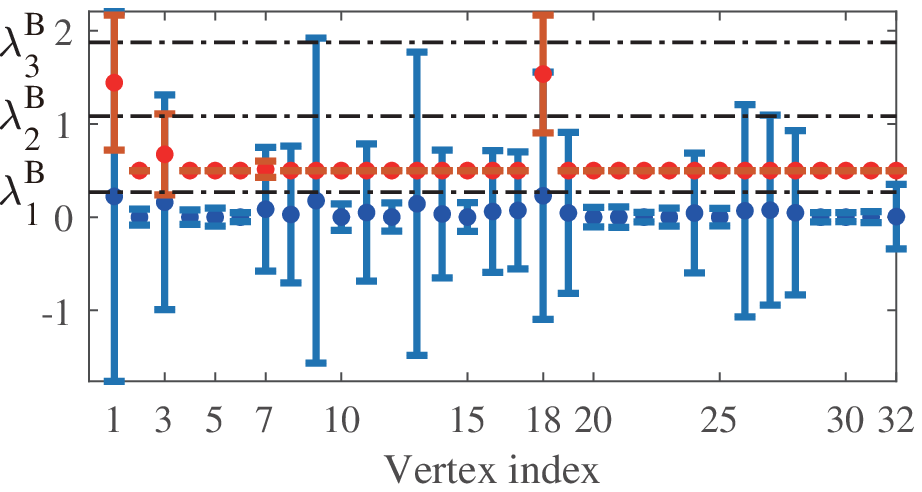}
  \end{minipage}
}
\caption{Centers and radiuses of Gershgorin discs.}
\label{GD}
\end{figure}

Next, the centers and radiuses of Gershgorin discs for non-Bayesian DoS and Bayesian DoS are shown in Fig. \ref{GD}(a) and Fig. \ref{GD}(b), respectively. 
Those with uniform sampling proportion are blue and those with designed sampling proportion are red. The dots and the bars represent the centers and radiuses.
We can find that the prior makes the 1st, 3rd and 18th center in blue go higher and the 7th center in blue goes lower in Fig. \ref{GD}(b) compared to Fig. \ref{GD}(a). 
For red centers, the design of the sampling proportion makes a further increase in this trend.

\begin{figure}[!t]
  \centering
\begin{minipage}[b]{\linewidth}
  \centering
  \includegraphics[scale = 0.38]{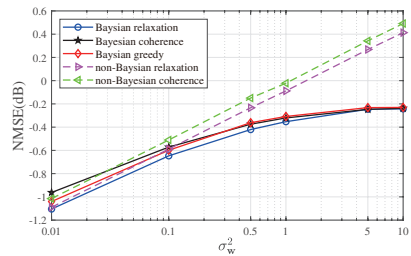}
\end{minipage}
\caption{NMSE for different algorithms versus different observation noise.}
\label{error}
\vspace{-0.5cm}
\end{figure}

Finally, we numerically evaluate the performance by the normalized MSE (NMSE) of different sample set selection algorithms versus different variance of the noise in Fig. \ref{error}.
We compare our approach with the following methods: non-Bayesian coherence \cite{puy2018random}, non-Bayesian relaxation \cite{xie2017design} and Bayesian greedy \cite{shamaiah2010greedy}. 
It can be seen that the performance of GS estimation can be improved when any vertex is allowed to be sampled for multiple times in both non-Bayesian and Bayesian DoS.
The performance of Bayesian DoS is better than non-Bayesian DoS especially when the variance of the noise is large. 
Our heuristic DoS based on Bayesian coherence leads to a comparable performance suggesting that it can be applied for large scale GS when we do not have enough computing power for solving the optimization problem.

\section{Conclusion}
In this paper, we proposed an optimization framework for Bayesian DoS to make the full use of the stochastic prior knowledge of bandlimited GS.
And we also proposed a heuristic algorithm for GS to reduce the calculation burden.
In future work, we are going to design more different utility functions for our framework to achieve different goals of Bayesian DoS, for example, prediction GS with the least error.

\section*{Acknowledgment}
This work was supported by the National Key Research and Development Program of China (No. 213), the Shanghai Municipal Natural Science Foundation (No. 19ZR1404700), and the NSF of China (No. 61501124).





%

\bibliographystyle{IEEEbib}
\bibliography{strings,refs}

\end{document}